\pdfoutput=1
\documentclass[useAMS,usenatbib]{mn2e}
\usepackage{graphicx}
\usepackage{amssymb}
\usepackage{lscape}
\usepackage{rotating}
\usepackage{subfigure}
\usepackage{amsmath}
\def\deg{\ifmmode^\circ\else$^\circ$\fi}

\def\Q{\ifmmode\mathcal{Q}\else$\mathcal{Q}$\fi}
\def\Mach{\ifmmode\mathcal{M}\else$\mathcal{M}$\fi}

\title[Uncovering physical environments around W33]
{Uncovering distinct environments in an extended physical system around the W33 complex}
\author[L.~K. Dewangan et al.]
{L.~K. Dewangan$^{1}$\thanks{lokeshd@prl.res.in}, T. Baug$^{2}$, and D.~K. Ojha$^{3}$\\
$^{1}$Physical Research Laboratory, Navrangpura, Ahmedabad - 380 009, India.\\
$^2$Kavli Institute for Astronomy and Astrophysics, Peking 
University, 5 Yiheyuan Road, Haidian District, Beijing 100871, P. R. China.\\
$^3$Department of Astronomy and Astrophysics, Tata Institute of Fundamental Research, Homi Bhabha Road, 
Mumbai 400 005, India.\\}

\begin{document}

\date{ }

\pagerange{\pageref{firstpage}--\pageref{lastpage}} \pubyear{2020}

\maketitle

\label{firstpage}

\begin{abstract}
We present a multi-wavelength investigation of a 
large-scale physical system containing the W33 complex. The extended system ($\sim$50 pc $\times$ 37 pc) is selected based on the 
distribution of molecular gas at [29.6, 60.2] km s$^{-1}$ and of 88 ATLASGAL 870 $\mu$m dust clumps at d $\sim$2.6 kpc.
The extended system/molecular cloud traced in the maps of $^{13}$CO and C$^{18}$O emission contains several H\,{\sc ii} regions excited by OB stars 
(age $\sim$0.3--1.0 Myr) and  a thermally supercritical filament (``fs1", length $\sim$17 pc).
The filament, devoid of the ionized gas, shows dust temperature (T$_{d}$) of $\sim$19~K, while the H\,{\sc ii} regions are depicted with T$_{d}$ of $\sim$21--29~K. 
It suggests the existence of two distinct environments in the cloud. 
The distribution of Class~I young stellar objects (mean age $\sim$0.44 Myr) traces the early stage of star formation (SF) toward the cloud. 
At least three velocity components (around 35, 45, and 53 km s$^{-1}$) are investigated toward the system. 
The analysis of $^{13}$CO and C$^{18}$O reveals the spatial and velocity connections of cloud components around 35 and 53 km s$^{-1}$. 
The observed positions of previously known sources, W33~Main, W33~A and O4-7I stars, are found toward a complementary distribution of these two cloud components.
The filament ``fs1" and a previously known object W33~B are seen toward the overlapping areas of the clouds, where ongoing SF activity is evident. 
A scenario concerning the converging/colliding flows from two different velocity components appears to explain well the observed signposts of SF activities in the system.
\end{abstract}
\section{Introduction}
\label{sec:intro}
Over the past two decades, due to the availability of various large-scale multi-wavelength 
surveys, the research concerning the understanding of the formation processes of massive 
OB stars ($\gtrsim$ 8 M$_{\odot}$) and young stellar clusters has received immense interest, 
and is being debated \citep{zinnecker07,tan14}. In this context, Galactic star-forming sites hosting 
mid-infrared (MIR) bubbles/shells associated with H\,{\sc ii} regions and/or infrared dark clouds (IRDCs)/elongated filamentary features 
are thought to be prime targets for exploring the highlighted research problem \citep[e.g.,][]{churchwell06,churchwell07,andre10,andre14}. The exploration of the existing multi-wavelength surveys allows to examine the physical conditions and the kinematics 
of embedded structures on large-scale areas of these sites, which hold crucial clues about the birth processes of clusters of young stellar objects (YSOs) 
and massive stars. In this connection, the target of this paper is a massive star-forming site W33, which is located in 
the inner Galaxy. Previous studies have reported that W33 hosts several massive OB stars and star-forming clumps, and filaments 
\citep[][and references therein]{westerhout58,immer13,immer14,maud17,kohno18}. 

Based on the radio line and continuum observations \citep{downes80,haschick83,lockman89,anderson14,anderson15} and 
the VLT/SINFONI near-infrared (NIR) spectroscopic observations \citep{messineo15}, several ionized clumps/H\,{\sc ii} 
regions and massive OB stars have been investigated in the W33 complex \citep[see Figure~1 in][]{kohno18}. 
In the W33 complex, \citet{messineo15} reported that the population of massive stars (including O4-7 type stars) formed during $\sim$2--4 Myr ago.
At least six dust continuum clumps at 870 $\mu$m were detected toward the W33 complex within an area of 15 pc $\times$ 15 pc \citep[e.g.,][]{contreras13}, which were referred to as 
W33 Main, W33 A, W33 B, W33 Main1, W33 A1, and W33 B1 \citep[see Figure~1 in][]{kohno18}. 
\citet{kohno18} identified two CO bipolar outflows toward W33 Main and W33 A, and the dynamical timescales of those outflows were computed to be $\sim$3 $\times$ 10$^{4}$ yr. 
\citet{immer13} detected water masers toward W33 A (G012.90$-$0.24, G012.90$-$0.26) and W33 Main (G012.81$-$0.19) in a radial velocity (V$_\mathrm{lsr}$) range of 33--38 km s$^{-1}$, 
while the water masers in W33 B (G012.68$-$0.18) were observed in a velocity range of 57--63  km s$^{-1}$. 
Note that the radial velocity information has been found different for W33 B compared to W33 A and W33 Main. These authors suggested the kinematic 
distance of $\sim$3.7 kpc (corresponding to V$_\mathrm{lsr}$ of 36 km s$^{-1}$). 
They also reported the parallactic distance of the water masers in the massive star forming complex W33 
(containing G012.68$-$0.18, G012.81$-$0.19, G012.90$-$0.24, G012.90$-$0.26) to be $\sim$2.4 kpc.

It was suggested that the distance estimated using the trigonometric parallax observations of maser sources is more reliable than the kinematic distance measurements \citep[e.g.,][]{reid09,sato10,immer13}. Hence, despite different V$_\mathrm{lsr}$ values, it has been considered that the sites (i.e., W33 A (G012.90$-$0.24, G012.90$-$0.26), W33 B (G012.68$-$0.18), and W33 Main (G012.81$-$0.19)) are located at the similar distance, and are part of the W33 complex \citep[e.g.,][]{immer13,messineo15,kohno18}. 
Based on the detection of methanol masers, \citet{kohno18} also pointed out the presence of massive YSOs toward the sub-regions W33 Main, W33 A, and W33 B. 
Using the high-resolution molecular line data, inner environments of at least two sub-regions W33 A and W33 Main have been 
reported in the literature \citep{galvan10,jiang15,maud17}. 
To explain star formation (SF) activity in the direction of W33 A, \citet{galvan10} proposed a triggered SF scenario 
by filamentary convergent gas flows from two different velocity components. 
In the direction of the W33 complex, \citet{kohno18} reported three velocity components at 35, 45, and 58 km s$^{-1}$, 
and found the signatures of the collision between two clouds at 35 and 58 km s$^{-1}$. 
They assumed these clouds at a distance of $\sim$2.4 kpc. 
Considering this argument, a collision scenario was proposed to explain the massive SF in the W33 complex \citep[see][for more details]{kohno18}. 
\citet{messineo15} also mentioned the possibility of sequential SF and feedback in W33. 
However, such study is not extensively carried out in W33. To our knowledge, earlier observational works toward W33 are restricted up to an area of 15 pc $\times$ 15 pc \citep[see Figure~1 in][]{kohno18}, 
and its large-scale environment (i.e., $>$ 35 pc $\times$ 35 pc) is yet to be examined. 

In this context, Figure~\ref{fig1}a displays a large-scale area ($\sim$1$\degr$.1 $\times$ 0$\degr$.815) containing the W33 complex using the 
ATLASGAL 870 $\mu$m dust continuum map. Figure~\ref{fig1}a also indicates an area ($\sim$0$\degr$.4 $\times$ 0$\degr$.4) by a broken box, which was studied by \citet{kohno18}. 
The locations of W33 A (G012.90$-$0.24, G012.90$-$0.26), W33 B (G012.68$-$0.18), and W33 Main (G012.81$-$0.19) are also marked by filled circles in Figure~\ref{fig1}a. 
In Figure~\ref{fig1}a, the positions of 88 ATLASGAL clumps at 870 $\mu$m \citep[from][]{urquhart18} are also marked by diamonds. 
\citet{urquhart18} obtained velocities toward the ATLASGAL clumps and also tried to compute the distance information for the clumps. 
All these ATLASGAL clumps are depicted in a velocity range of [30, 56] km s$^{-1}$, and 
are located at a distance of $\sim$2.6 kpc \citep{urquhart18}, which are in agreement with the previously published results \citep[e.g.,][]{immer13,kohno18}. 
In this paper, we have adopted a distance of $\sim$2.6 kpc to the entire selected area around the W33 complex for all the analysis. 

One can also notice that the ATLASGAL survey has allowed us to trace the boundary of an extended physical system hosting the W33 complex (area $\sim$50 pc $\times$ 37 pc; distance $\sim$2.6 kpc; see Figure~\ref{fig1}a). However, we do not find any study related to unearth the physical conditions and the kinematics 
of embedded structures on large-scale areas. 
In this paper, using a multi-wavelength approach, we aim to explore the ongoing physical processes in the extended physical system hosting W33. In particular, this observational work focuses 
to understand the formation mechanisms of massive stars and clusters of YSOs. In this connection, the study of molecular gas toward our selected extended system has been carefully carried out using the FOREST Unbiased Galactic plane Imaging survey with the Nobeyama 45-m 
telescope \citep[FUGIN;][]{umemoto17} $^{13}$CO(J = 1--0) and C$^{18}$O (J = 1--0) line data. 

The present paper is organized as follows. The information of the adopted data sets is given in Section~\ref{sec:obser}. 
In Section~\ref{sec:data}, we present the observational findings derived using the multi-wavelength surveys. 
We discuss the possible SF processes in Section~\ref{sec:disc}. 
Finally, the main conclusions of this paper are presented in Section~\ref{sec:conc}.
\section{Data sets and analysis}
\label{sec:obser}
With the aid of the spatial distribution of 88 ATLASGAL clumps at a distance of $\sim$2.6 kpc, 
the selected target area ($\sim$1$\degr$.1 $\times$ 0$\degr$.815 (or $\sim$50 pc $\times$ 37 pc); centered at $l$ = 12$\degr$.946; $b$ = $-$0$\degr$.192) around W33 is presented in Figure~\ref{fig1}a. 
Table~\ref{ftab1} gives a summary of various multi-wavelength survey data adopted in this paper.
In the selected wide-scale area, these surveys offer to examine the distribution of H$_{2}$ column densities, dust temperatures, dust clumps, 
ionized gas, YSOs, and gas kinematics as well as the embedded structures/morphologies. 

In order to study the gas distribution toward the selected target area, the present paper employs the molecular $^{13}$CO (J=1--0) 
and C$^{18}$O (J=1--0) line data from the FUGIN survey \citep{umemoto17}. In the survey, the multi-beam receiver, FOur-beam REceiver System on the 45-m Telescope \citep[FOREST;][]{minamidani16,nakajima19}, was used for the observations. The FUGIN molecular line data are calibrated in main beam temperature ($T_\mathrm{mb}$) \citep[see][for more details]{umemoto17}. 
The typical rms noise level\footnote[1]{https://nro-fugin.github.io/status/} ($T_\mathrm{mb}$) is $\sim$0.7~K for both the $^{13}$CO and C$^{18}$O lines \citep{umemoto17}. 
To improve sensitivities, we smoothed the FUGIN data cube with a Gaussian function having a half power beam width of 35$''$. 
 \begin{table*}
 % \tiny
\setlength{\tabcolsep}{0.05in}
\centering
\caption{Details of different surveys utilized in this paper.}
\label{ftab1}
\begin{tabular}{lcccr}
\hline 
  Survey  &  band/line(s)       &  Resolution ($\arcsec$)        &  Reference \\   
\hline
\hline 
% Multi-Array Galactic Plane Imaging Survey (MAGPIS)                             & 20 cm                       & $\sim$6          & \citet{helfand06}\\
% The HI/OH/Recombination line survey of the inner Milky Way (THOR)                             & 1--2 GHz                       & $\sim$25          & \citet{beuther16}\\
 NRAO VLA Sky Survey (NVSS)                                   & 21 cm                       & $\sim$46          & \citet{condon98}\\
 FUGIN survey  &   $^{13}$CO, C$^{18}$O (J = 1--0) & $\sim$20        &\citet{umemoto17}\\
% Galactic Ring Survey (GRS)                                                                   & $^{13}$CO (J = 1--0) & $\sim$45        &\citet{jackson06}\\
APEX Telescope Large Area Survey of the Galaxy (ATLASGAL)                 &870 $\mu$m                     & $\sim$19.2        &\citet{schuller09}\\
{\it Herschel} Infrared Galactic Plane Survey (Hi-GAL)                              &70--500 $\mu$m                     & $\sim$5.8--37         &\citet{molinari10}\\
{\it Spitzer} MIPS Inner Galactic Plane Survey (MIPSGAL)                                         &24 $\mu$m                     & $\sim$6         &\citet{carey05}\\ 
{\it Spitzer} Galactic Legacy Infrared Mid-Plane Survey Extraordinaire (GLIMPSE)       &3.6--8.0  $\mu$m                   & $\sim$2           &\citet{benjamin03}\\
%UKIRT near-infrared Galactic Plane Survey (GPS)                                                 &1.25--2.2 $\mu$m                   &$\sim$0.8           &\citet{lawrence07}\\ 
%Two Micron All Sky Survey (2MASS)                                                 &1.25--2.2 $\mu$m                  & $\sim$2.5          &\citet{skrutskie06}\\
\hline          
\end{tabular}
\end{table*}
\section{Results}
\label{sec:data}
\subsection{Wide-scale view around W33: environment and H\,{\sc ii} regions}
\label{subsec:radio}
\begin{figure*}
\includegraphics[width=9.2cm]{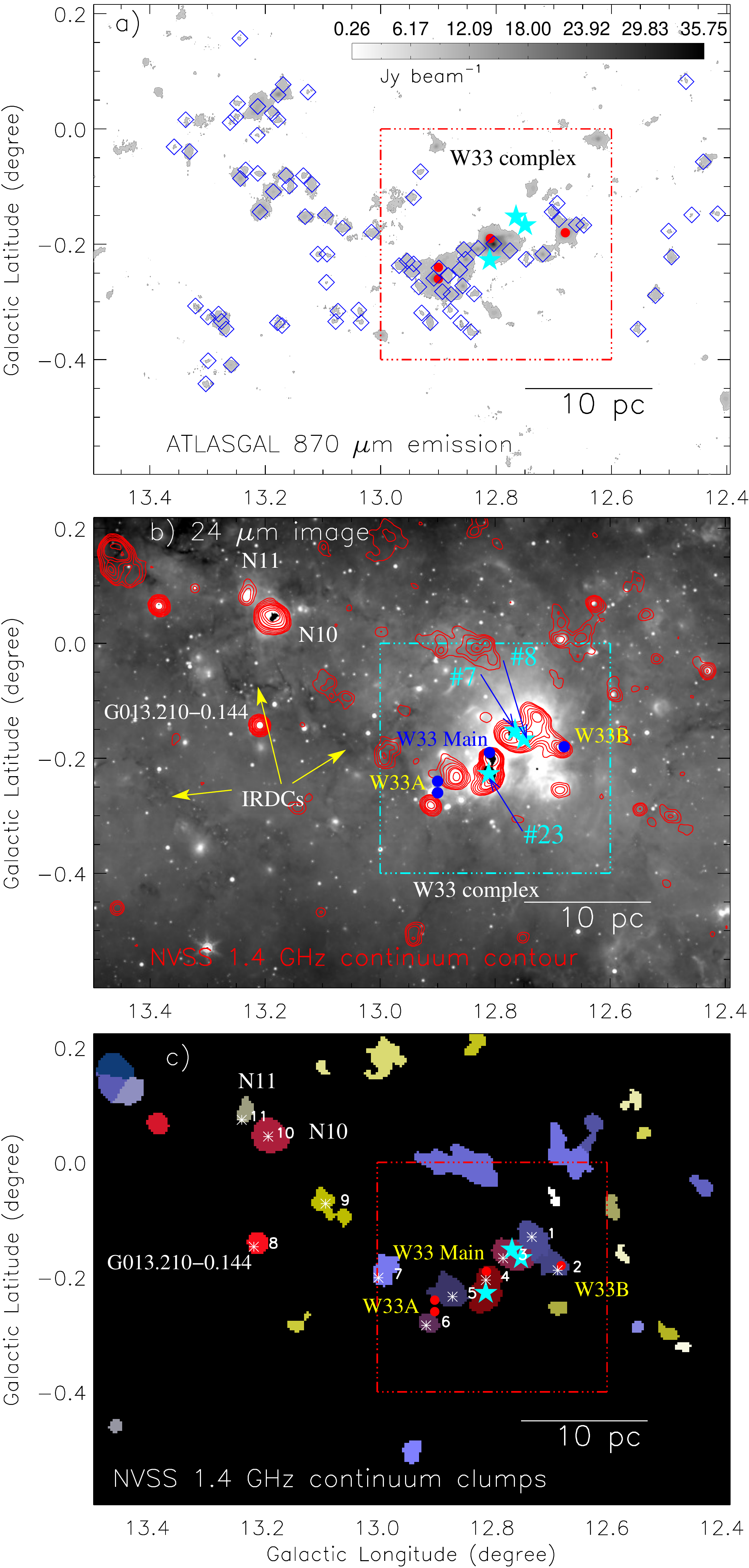}
\caption{a) Overlay of 88 ATLASGAL dust continuum clumps at 870 $\mu$m \citep[from][]{urquhart18} on the ATLASGAL 
contour map at 870 $\mu$m (area $\sim$1$\degr$.1 $\times$ 0$\degr$.815 
($\sim$50 pc $\times$ 37 pc at a distance of 2.6 kpc); central coordinates: $l$ = 12$\degr$.946; $b$ = $-$0$\degr$.192).
The ATLASGAL contour levels are 36.48 Jy/beam $\times$ (0.0072, 0.012, 0.03, 0.05, 0.067, 0.08, 0.1, 0.15, 0.2, 0.25, 0.3, 0.4, 0.5, 0.6, 0.7, 0.8, 0.9, 0.95, and 0.98). 
The ATLASGAL clumps (at d $\sim$2.6 kpc) are shown by blue diamonds. 
b) Overlay of the NVSS 1.4 GHz continuum emission contours (in red) on the MIPSGAL 24 $\mu$m image. 
The NVSS radio continuum contour levels are 4.7, 9.4, 18.8, 37.7, 62.8, 85.0, 156.9, 251.0, 367.5, and  1255.10 mJy beam$^{-1}$. 
The locations of G013.210$-$0.144, bubbles N10 and N11 as well as IRDCs are indicated in the figure. 
c) The boundary of selected radio clump in the NVSS 1.4 GHz radio continuum map is shown along with its corresponding ID and position (see asterisks and also Table~\ref{tab2}). 
In each panel, filled stars (in cyan) show the locations of O4-7 type stars \citep[from][]{messineo15} and filled circles highlight some sub-regions in the W33 complex (i.e., W33 A (G012.90$-$0.24, G012.90$-$0.26), W33 B (G012.68$-$0.18), and W33 Main (G012.81$-$0.19)). In all the panels, the scale bar refers to 10 pc, and 
a broken box covering the field around the W33 complex presents the area investigated by \citet{kohno18}.} 
\label{fig1}
\end{figure*}

Figure~\ref{fig1}a exhibits the spatial distribution of 88 ATLASGAL clumps (having d $\sim$2.6 kpc 
and V$_\mathrm{lsr}$ range $\sim$[30, 56] km s$^{-1}$) overlaid on the ATLASGAL continuum map at 870 $\mu$m, illustrating an extended 
physical system as mentioned earlier. Figure~\ref{fig1}b shows the overlay of the NVSS 
radio continuum emission contours on the MIPSGAL 24 $\mu$m image, revealing the 
locations of H\,{\sc ii} regions in our selected system. 
The NVSS 1.4 GHz radio continuum data suggest the presence of the ionized gas, while the warm dust emission is traced in the 24 $\mu$m image. 
Additionally, one can also find the absorption features against the Galactic background in the 
24 $\mu$m image, revealing the IRDCs (see arrows in Figure~\ref{fig1}b). 
The locations of previously identified at least three O4-7(super)-giant stars \citep[see IDs \#7, 8, and 23; from][]{messineo15} are 
marked in Figures~\ref{fig1}a and~\ref{fig1}b (see filled stars). 
With the application of the {\it clumpfind} IDL program \citep{williams94} in the NVSS 1.4 GHz radio continuum map, we have selected 11 ionized clumps in our target area. 
Figure~\ref{fig1}c shows the boundary of each NVSS clump against the locations of earlier reported O4-7(super)-giant stars (see filled stars). 
A broken box is indicated in Figures~\ref{fig1}a,~\ref{fig1}b and~\ref{fig1}c, which shows the area investigated by \citet{kohno18}. 
The ionized clumps (see IDs \#c10, \#c8, and \#c1--c7) are found toward the MIR bubbles \citep[i.e., N10 and N11;][]{churchwell06,gama16}, 
G013.210$-$0.144 \citep{white05}, and the W33 complex (W33 A (G012.90$-$0.24, G012.90$-$0.26), W33 B (G012.68$-$0.18), 
and W33 Main (G012.81$-$0.19); see Figure~\ref{fig1}c), where we also find extended warm dust emission at 24 $\mu$m (see Figure~\ref{fig1}b). 
Two massive O4-7(super)-giant stars (i.e., \#7 and \#8) are seen toward the NVSS clump \#c3, while a massive O4-7(super)-giant star \#23 is 
found in the direction of the NVSS clump \#c4 (see Figure~\ref{fig1}c). 

Following the work of \citet{dewangan17}, we have estimated the number of Lyman continuum photons \citep[$N_\mathrm{uv}$; see][]{panagia73,matsakis76} and 
the dynamical age \citep[$t_\mathrm{dyn}$; see][]{dyson80} of each ionized clump (see Table~\ref{tab2}). 
Previously, \citet{kohno18} adopted an initial particle number density of the ambient 
neutral gas (i.e., $n_{0}$ = 10$^{4}$ cm$^{-3}$) in the estimations of the formation timescale of the H\,{\sc ii} regions G012.745$-$00.153 and G012.820$-$00.238, which was considered as a reasonable value for the W33 complex. Following the previous work of \citet{kohno18}, in this paper, we compute the ages of the ionized clumps for a value of $n_{0}$ = 10$^{4}$ cm$^{-3}$. 
The calculation uses the isothermal sound velocity in the ionized gas \citep[$c_{s}$ = 11 km s$^{-1}$;][]{bisbas09}, 
the radius of the H\,{\sc ii} region ($R_{HII}$), $n_{0}$, and $N_\mathrm{uv}$. 
Dynamical ages of the ionized clumps vary between $\sim$0.35 -- 1.0 Myr (see Table~\ref{tab2}). 
The bubble N10 (see ID \#c10 in Table~\ref{tab2}) is found to be excited by an O9.5V--O9V star (age $\sim$0.65 Myr), which is located away from the W33 complex. 
At least seven NVSS clumps (see IDs \#c1--c7) are located toward the W33 complex (area $\sim$15 pc $\times$ 15 pc; see a broken box in Figure~\ref{fig1}c), which contains OB 
stars (age $\sim$0.35 -- 0.85 Myr). More description of the analysis can be found in \citet{dewangan17}. 
The implication of the ages of these ionized clumps is discussed in Section~\ref{sec:disc}.
\subsection{Dust temperature and column density maps}
\label{sec:hermap} 
In order to study the embedded structures in our selected target area, we have obtained the {\it Herschel} temperature 
and column density ($N(\mathrm H_2)$) maps from the site\footnote[2]{http://www.astro.cardiff.ac.uk/research/ViaLactea/}. 
The spatial resolution of these maps is $\sim$12$''$. Using the Bayesian {\it PPMAP} procedure operated on the {\it Herschel} 
data at wavelengths of 70, 160, 250, 350 and 500 $\mu$m \citep{marsh15,marsh17}, the {\it Herschel} temperature and column density 
maps were produced for the {\it EU-funded ViaLactea project} \citep{molinari10b}. 

\begin{figure*}
\includegraphics[width=13.4cm]{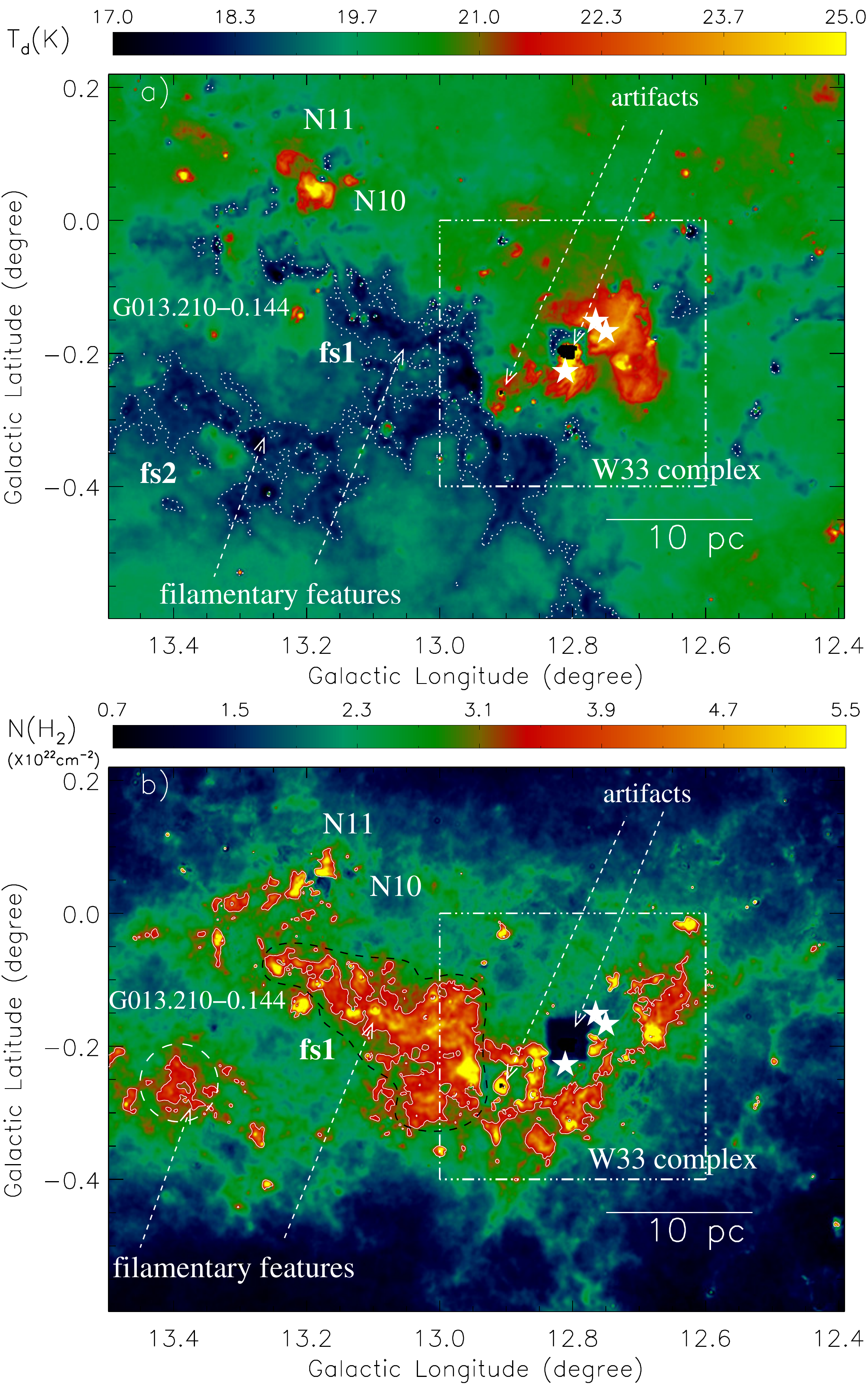}
\caption{a) The panel shows the {\it Herschel} temperature map of the selected area around W33. 
A broken contour (in white) shows the temperature feature at 18.6 K in the figure, tracing at least two filamentary features (i.e., ``fs1" and ``fs2"). 
b) The panel shows the {\it Herschel} column density ($N(\mathrm H_2)$) map of the selected area. 
The column density contour (in white) is also overlaid on the map with a level of 3.35 $\times$ 10$^{22}$ cm$^{-2}$. 
The filament ``fs1" is indicated by a broken curve (in black). In each panel, the scale bar refers to 10 pc, and 
a broken box covering the field around the W33 complex presents the area investigated by \citet{kohno18}. 
In both the panels, artifacts are also indicated by arrows, and filled stars (in white) show the locations of O4-7 type stars (see Figure~\ref{fig1}b).}
\label{fig2}
\end{figure*}

In Figures~\ref{fig2}a and~\ref{fig2}b, we display the {\it Herschel} temperature and column density ($N(\mathrm H_2)$) maps of our selected target area, respectively. 
In both the {\it Herschel} maps, a broken box shows the area studied by \citet{kohno18}. 
The {\it Herschel} temperature map shows the existence of embedded structures with T$_\mathrm{d}$ $\sim$17--19~K in the system, and is overlaid with a 
temperature contour of 18.6~K, allowing us to depict filamentary structures (see labels fs1 and fs2 in Figure~\ref{fig2}a). 
The {\it Herschel} temperature map also exhibits extended features with T$_\mathrm{d}$ $\sim$21--29~K in the direction of the previously 
known H\,{\sc ii} regions (see W33~A (G012.90$-$0.24, G012.90$-$0.26), W33~B (G012.68$-$0.18), W33~Main (G012.81$-$0.19), MIR bubbles, and G013.210$-$0.144). 

In Figure~\ref{fig2}b, the column density contour (in white) is also overlaid on the $N(\mathrm H_2)$ map with a level of 3.35 $\times$ 10$^{22}$ cm$^{-2}$, 
indicating the presence of materials with high column densities. 
Using this $N(\mathrm H_2)$ contour level, we trace the boundary of the filament ``fs1" (length $\sim$17 pc), which is also highlighted by a broken black 
curve in Figure~\ref{fig2}b. However, an elongated morphology of filament ``fs2", as traced in the {\it Herschel} temperature map, is not seen by this $N(\mathrm H_2)$ contour level. One of the parts of the filament ``fs2" is prominently depicted by this $N(\mathrm H_2)$ contour level (see a broken white circle). 
In the direction of the IRDCs highlighted in the image at 24 $\mu$m, both the {\it Herschel} maps indicate the presence of the embedded filaments. 
With the knowledge of T$_\mathrm{d}$ toward the filamentary structures and the H\,{\sc ii} regions, two distinct environments are evident 
in the selected physical system. 

Based on the analysis of the {\it Herschel} column density and temperature maps, we examine the stability of the elongated 
filament ``fs1" (mass $\sim$51000~M$_{\odot}$; length $\sim$17~pc; T$_\mathrm{d}$ $\sim$19~K). 
To compute the mass of the filament/clump, we employed the equation, $M_{area} = \mu_{H_2} m_H Area_{pix} \Sigma N(H_2)$, where $\mu_{H_2}$ is 
the mean molecular weight per hydrogen molecule (i.e., 2.8), $Area_{pix}$ is the area subtended by one pixel (i.e., 6$''$/pixel), and 
$\Sigma N(\mathrm H_2)$ is the total column density \citep[see also][]{dewangan17}. 
Following the works of \citet{dewangan18}, we have estimated the line mass or mass per unit length (i.e., $M_{\rm line,obs}$ $\sim$3000 M$_{\odot}$ pc$^{-1}$) and 
the critical line mass ($M_{\rm line,crit}$) of the filament ``fs1". 
Due to unknown inclination angle of the filament, we consider the observed line mass as an upper limit.
The expression of $M_{\rm line,crit}$ is given by $\sim$16~M$_{\odot}$ pc$^{-1}$ $\times$ (T$_{gas}$/10 K) for 
a gas filament, assuming that the filament is an infinitely extended, self-gravitating and 
isothermal cylinder without magnetic support \citep[e.g.][]{ostriker64,inutsuka97,andre14}. 
The value of $M_{\rm line,obs}$ can be compared against the critical line mass $M_{\rm line,crit}$ of 16--32 M$_{\odot}$ pc$^{-1}$ at T = 10--20 K, 
suggesting that the filament ``fs1" is thermally supercritical. In general, the thermally supercritical filaments are believed to be unstable to 
radial gravitational collapse and fragmentation \citep[e.g.,][]{andre10}.
\subsection{Star Formation Activities}
\label{sec:sf}
In this section, to examine the ongoing SF activities, we have identified Class~I YSOs in our selected target area around the W33 complex. 
In this context, the color-color plot ([4.5]$-$[5.8] vs [3.6]$-$[4.5]) is employed (not shown here), and the infrared color 
conditions (i.e., [4.5]$-$[5.8] $\ge$ 0.7 mag and [3.6]$-$[4.5] $\ge$ 0.7 mag) are adopted to depict Class~I YSOs \citep[see][for more details]{hartmann05,getman07,dewangan17b,dewangan18b,dewangan18x}. 
Photometric magnitudes of point-like sources at 3.6--5.8 $\mu$m were downloaded from the {\it Spitzer} GLIMPSE-I Spring' 07 highly reliable catalog. 
In this work, we considered only sources with a photometric error of less than 0.2 mag in each {\it Spitzer} band. 

\begin{figure*}
\includegraphics[width=10.2cm]{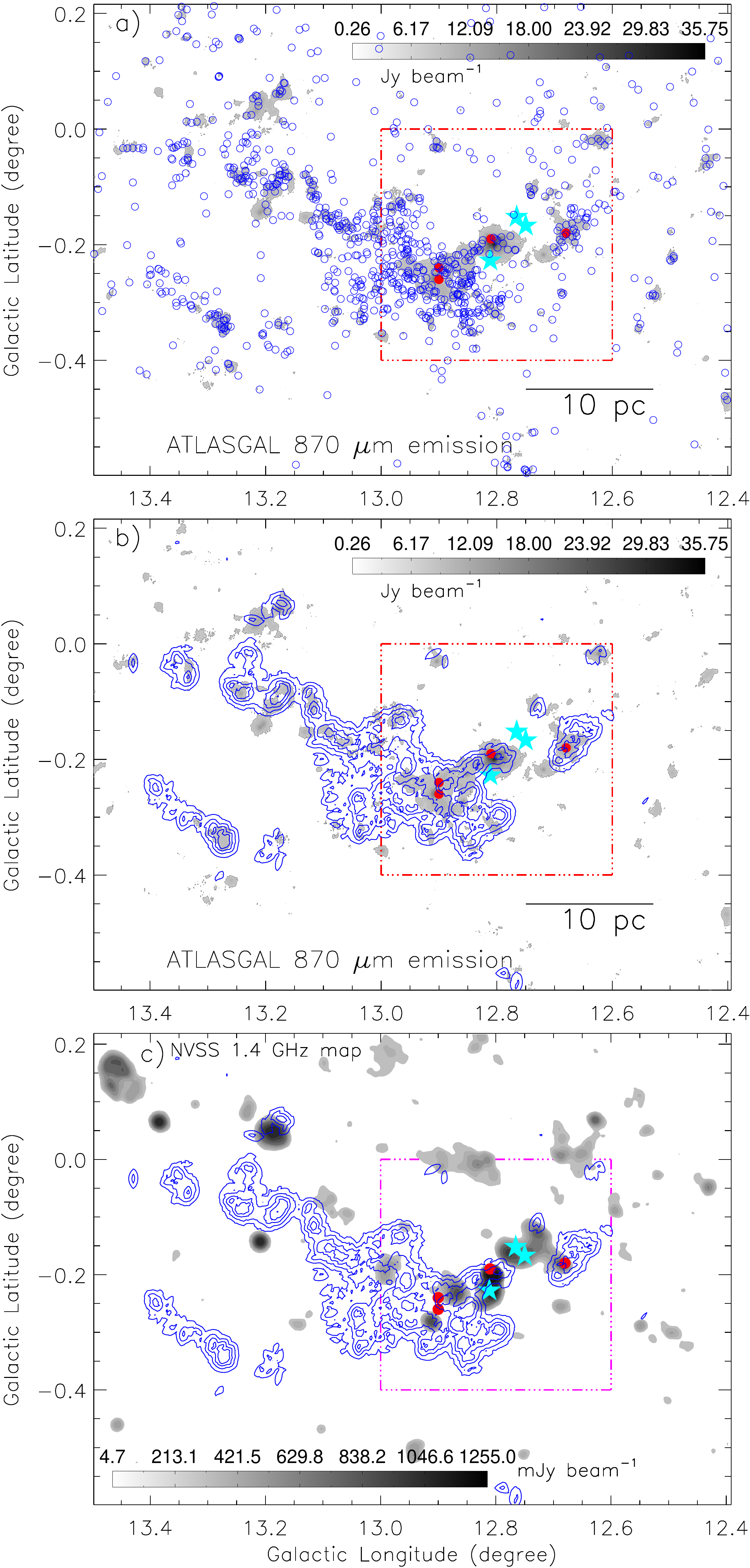}
\caption{a) Overlay of the positions of the selected Class~I YSOs on the ATLASGAL contour map at 870 $\mu$m. 
These YSOs (see open circles) satisfy the color conditions, [4.5]$-$[5.8] $\ge$ 0.7 and [3.6]$-$[4.5] $\ge$ 0.7 (see text for more details). 
b) The panel shows the overlay of the surface density contours (in blue) of YSOs on the ATLASGAL contour map at 870 $\mu$m.
The surface density contours (in cyan) of Class~I YSOs are shown with the levels of 1.2, 2, 3.5, and 6.5 YSOs/pc$^{2}$. 
c) Overlay of the surface density contours (in blue) on the NVSS 1.4 GHz continuum contour map. 
The ATLASGAL continuum map and the NVSS 1.4 GHz emission map are the same as in Figures~\ref{fig1}a and~\ref{fig1}b, respectively. 
In each panel, the other symbols (i.e., filled stars and filled circles) are the same as in Figure~\ref{fig1}b.}
\label{fig3}
\end{figure*}

Figure~\ref{fig3}a shows the overlay of the positions of 901 Class~I YSOs on the ATLASGAL 870 $\mu$m map (see blue circles). 
In Figures~\ref{fig3}b and~\ref{fig3}c, we present surface density contours (in blue) of Class~I YSOs overlaid on the ATLASGAL 870 $\mu$m dust continuum map 
and the NVSS 1.4 GHz radio continuum map, respectively. The surface density contours are shown with the levels of 1.2, 2, 3.5, and 6.5 YSOs/pc$^{2}$, where 1$\sigma$ = 1.1 YSOs pc$^{-2}$. 
A spatial correlation between the dust continuum emission and the Class~I YSOs is also clearly seen in Figure~\ref{fig3}b. 
A majority of Class~I YSOs \citep[mean age $\sim$0.44 Myr;][]{evans09} are 
seen toward the {\it Herschel} filaments (i.e.,``fs1" and ``fs2") and the H\,{\sc ii} regions, indicating the presence 
of the early stage of SF in the selected physical system. 
The surface density analysis is performed using the nearest-neighbour (NN) method \citep[see][for more details]{casertano85,gutermuth09,bressert10,dewangan17}. 
Following the similar procedures as reported in \citet{dewangan17}, we produce the surface density map of all the selected Class~I YSOs using a 15$''$ grid and 6~NN at a distance of 2.6~kpc. 
\subsection{Kinematics of molecular gas}
\label{sec:coem} 
\begin{figure*}
\includegraphics[width=13.2cm]{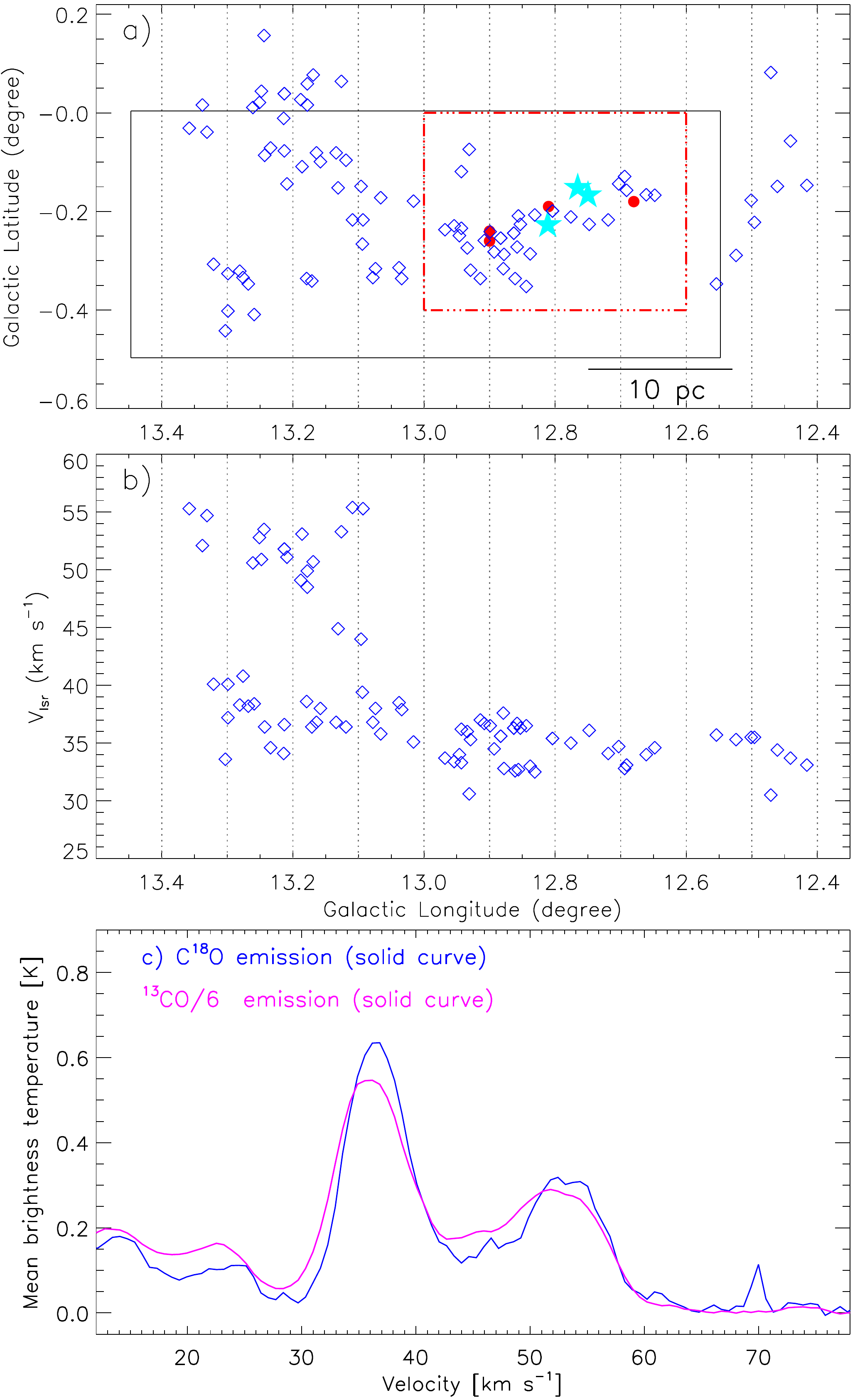}
\caption{a) The panel displays the spatial distribution of the ATLASGAL 870 $\mu$m dust continuum clumps (see diamonds) toward 
the selected site (see also Figure~\ref{fig1}a). Filled stars (in cyan) show the locations of O4-7 type stars, and filled circles (in red) highlight 
some sub-regions in the W33 complex (see Figure~\ref{fig1}b). b) The panel shows the distribution of the radial velocity against the Galactic longitude (see text for more details). 
c) The panel shows the FUGIN $^{13}$CO spectrum (see magenta curve) and the FUGIN C$^{18}$O spectrum (see blue curve). 
The profiles are produced by averaging the area highlighted by a solid box in Figure~\ref{fig4}a. 
The FUGIN $^{13}$CO spectrum has been divided by a factor of 6.}
\label{fig4}
\end{figure*}

In Figure~\ref{fig4}a, we display the positions of 88 ATLASGAL clumps (having d $\sim$2.6 kpc and V$_\mathrm{lsr}$ range $\sim$[30, 56] km s$^{-1}$). 
Figure~\ref{fig4}b presents the distribution of V$_\mathrm{lsr}$ of 88 clumps against the Galactic longitude. 
In the direction of {\it l} = 13$\degr$.08--13$\degr$.4, this plot suggests the presence of at least two 
clouds (at [30, 41] km s$^{-1}$ (or around 35 km s$^{-1}$) and [48, 56] km s$^{-1}$ (or around 53 km s$^{-1}$)) containing the dust clumps. 
Furthermore, we also find two clumps with V$_{lsr}$ of $\sim$45 km s$^{-1}$ toward {\it l} $\approx$ 13$\degr$.1. 
A majority of the ATLASGAL dust clumps are associated with the gas at [30, 41] km s$^{-1}$. 
In the direction of our selected system, our analysis of the clumps indicates the presence of at least three velocity components.
Previously, three cloud components at 35, 45, and 58 km s$^{-1}$ were also reported toward the W33 complex \citep[see][]{kohno18}. 
In this section, we have analyzed the FUGIN molecular line data to further explore various cloud components. 
In Figure~\ref{fig4}c, we display the observed $^{13}$CO(J =1$-$0) and C$^{18}$O(J =1$-$0) spectra in the direction of an area highlighted by a solid box in Figure~\ref{fig4}a), revealing two noticeable velocity peaks (i.e., 35 and 53 km s$^{-1}$). These profiles are produced by averaging the selected area.

\begin{figure*} 
\includegraphics[width=13.2cm]{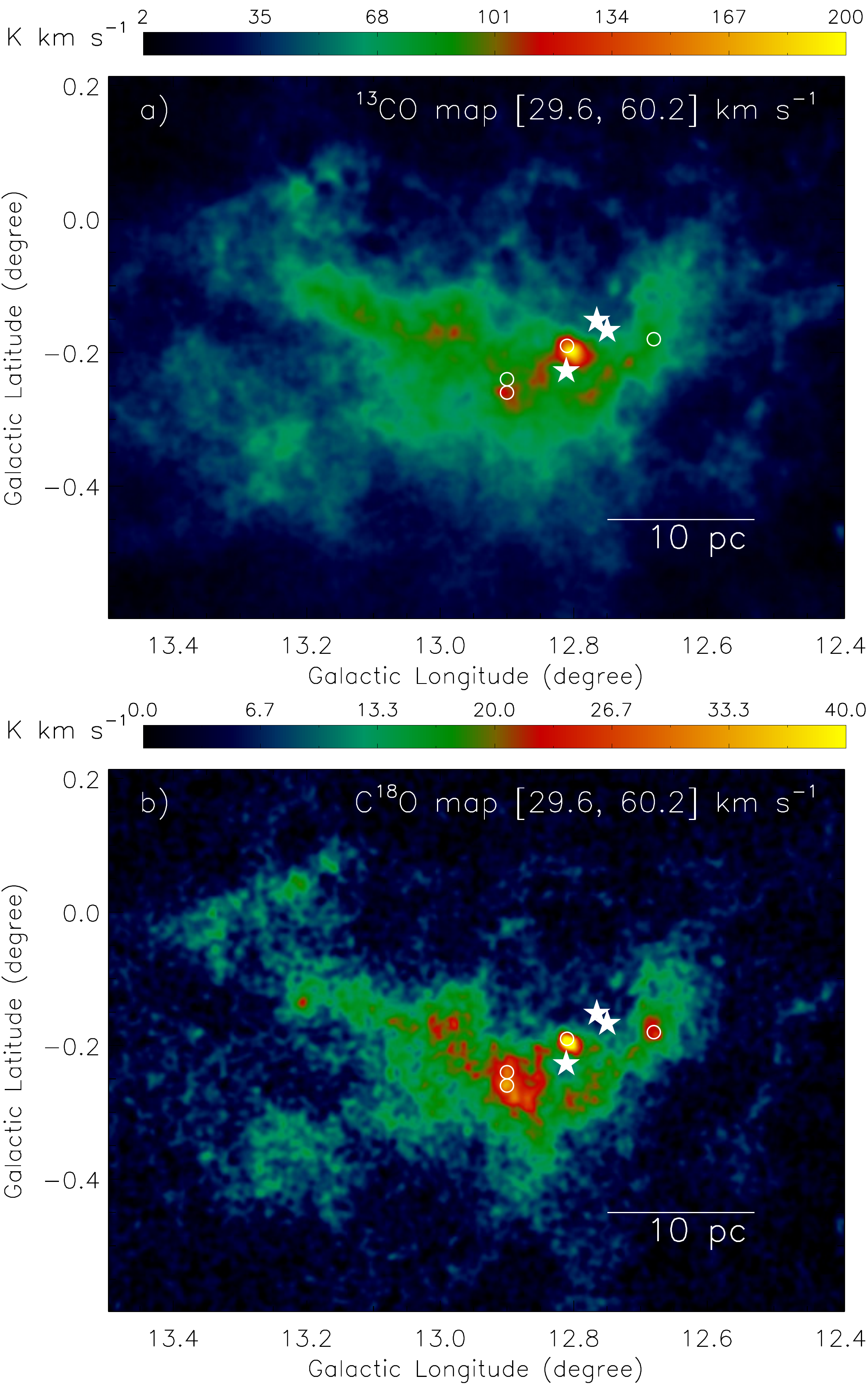}
\caption{a) FUGIN $^{13}$CO(J =1$-$0) map of intensity (moment-0) in the direction of the selected area around W33. 
b) FUGIN C$^{18}$O(J =1$-$0) map of intensity (moment-0). 
In each panel, the molecular emission is integrated from 29.6 to 60.2 km s$^{-1}$, 
and a scale bar corresponding to 10 pc is shown. 
In each panel, filled stars (in white) show the locations of O4-7 type stars and open circles (in white) highlight some sub-regions in the W33 complex (see Figure~\ref{fig1}b).} 
\label{fig5}
\end{figure*}

In the direction of the selected target area, Figures~\ref{fig5}a and~\ref{fig5}b display the integrated $^{13}$CO (J=1--0) 
and C$^{18}$O (J=1--0) intensity maps, respectively. 
In each intensity map, the molecular gas is integrated over a velocity range of [29.6, 60.2] km s$^{-1}$. 
Both the maps display an extended morphology of the molecular cloud associated with the selected target area, appearing like a giant molecular cloud (GMC). . 

\begin{figure*}
\centering
\includegraphics[width=17.0cm]{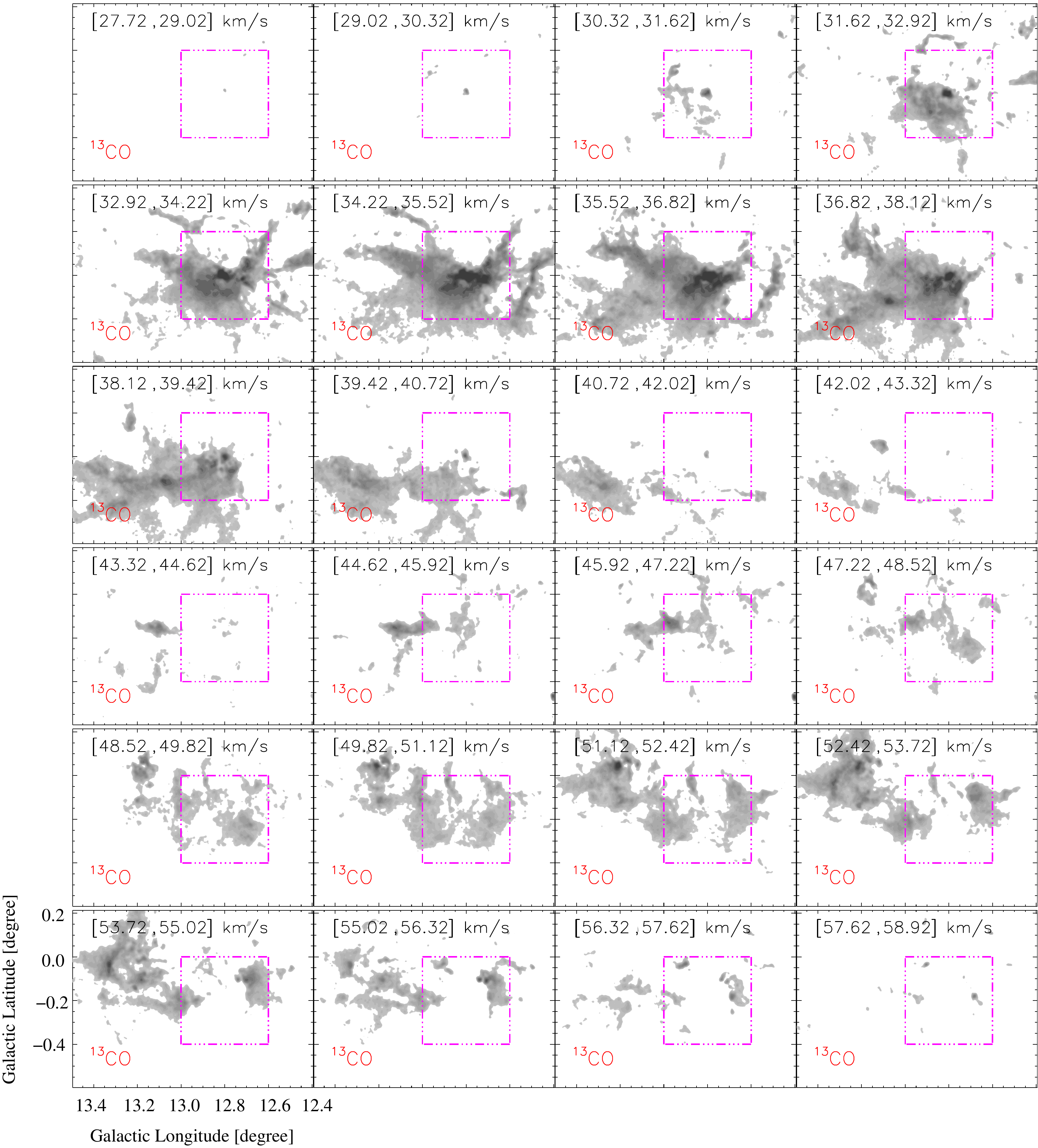}
\caption{The integrated velocity channel maps of $^{13}$CO(J =1$-$0) (at velocity intervals of 1.3 km s$^{-1}$). 
The $^{13}$CO contours are shown with the levels of 4.5, 5, 6, 7, 8, 9, 10, 11, 
13, 15, 18, 21, and 23 K km s$^{-1}$. In each panel a broken box, covering the field around the W33 complex, presents the area studied by \citet{kohno18}.\label{fig6}}
\end{figure*}

\begin{figure*} 
\includegraphics[width=16.2cm]{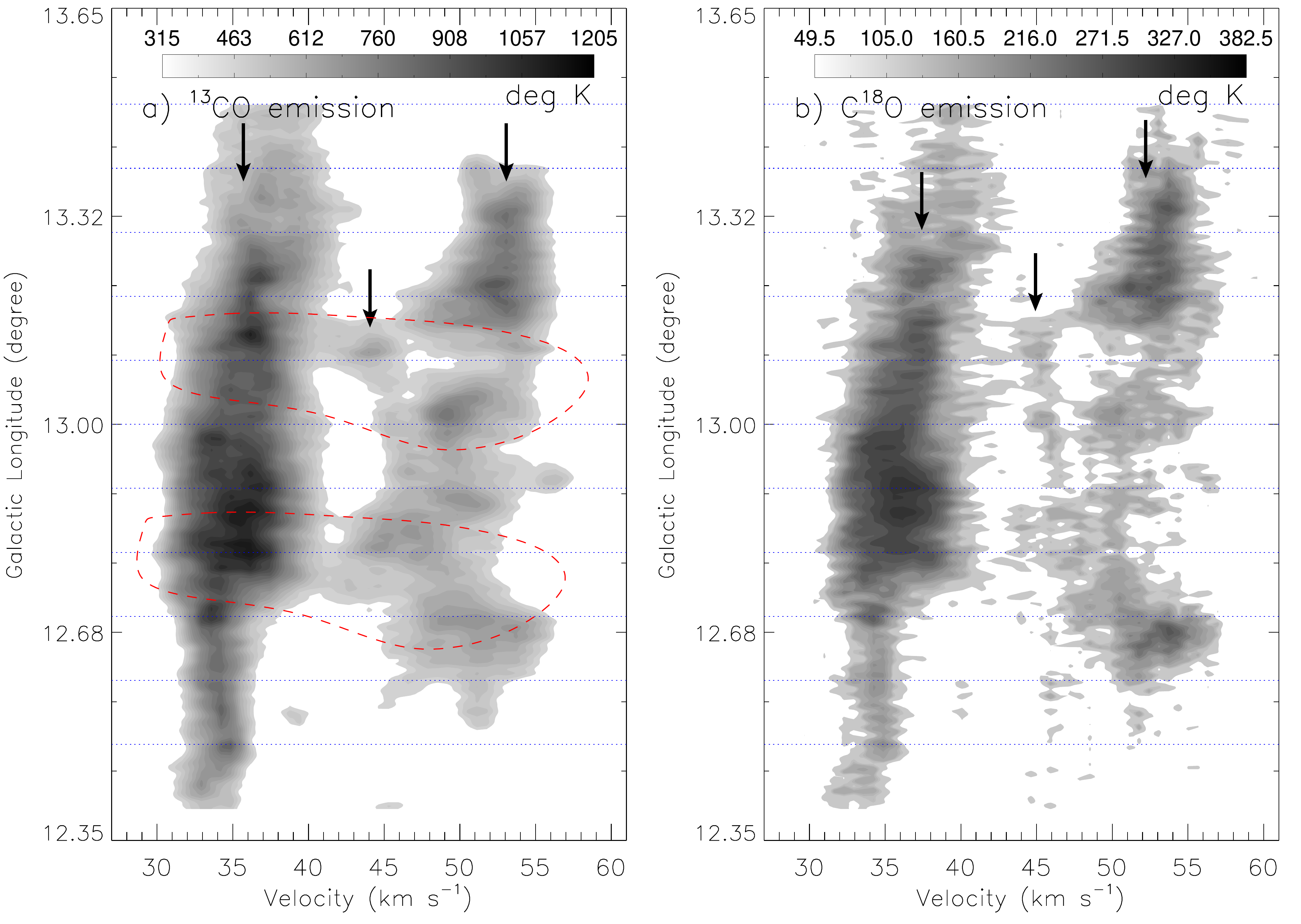}
\caption{a) Longitude-velocity map of $^{13}$CO. 
The contour levels are 0.45 deg K $\times$ (700, 800, 900, 1000, 1100, 1200, 1300, 1400, 1500, 1600, 1700, 1800, 1900, 2000, 2100, 2200, 2300, 2400, 2500, 2600, and 2680). Two broken curves (in red) show the connection of different cloud components. b) Longitude-velocity map of C$^{18}$O. The contour levels are 0.45 deg K $\times$ (110, 150, 160, 170, 210, 240, 270, 300, 330, 360, 400, 440, 500, 600, 700, 800, and 850).
Arrows indicate three velocity components in the direction of the selected area around W33. 
In each panel, the molecular emission is integrated over the latitude range from $-$0$\degr$.6 to 0$\degr$.21.}
\label{fig7}
\end{figure*}

Figure~\ref{fig6} presents the integrated $^{13}$CO velocity channel maps (at intervals of 1.3 km s$^{-1}$). 
The distribution of $^{13}$CO gas in the channel maps suggests the existence of different velocity components in the direction of 
the selected physical system (see panels at [36.82, 38.12], [44.62, 45.92], and [53.72, 55.02] km s$^{-1}$). 
In Figures~\ref{fig7}a and~\ref{fig7}b, we show the longitude-velocity maps of $^{13}$CO and C$^{18}$O, respectively. 
Both the position-velocity maps depict three velocity components around 35, 45, and 53 km s$^{-1}$ (see arrows in Figures~\ref{fig7}a and~\ref{fig7}b). 
Similar result is also derived using the distribution of the radial velocities 
of the ATLASGAL clumps (see Figure~\ref{fig4}b). 
In the velocity space, the cloud components around 35 and 53 km s$^{-1}$ appear to be connected by a low-intensity intermediate velocity 
emission around 45 km s$^{-1}$ (see red broken curves toward {\it l} = 13$\degr$.1--13$\degr$.2 and  {\it l} = 12$\degr$.7--12$\degr$.8 in Figure~\ref{fig7}a). 

\begin{figure*} 
\includegraphics[width=17.2cm]{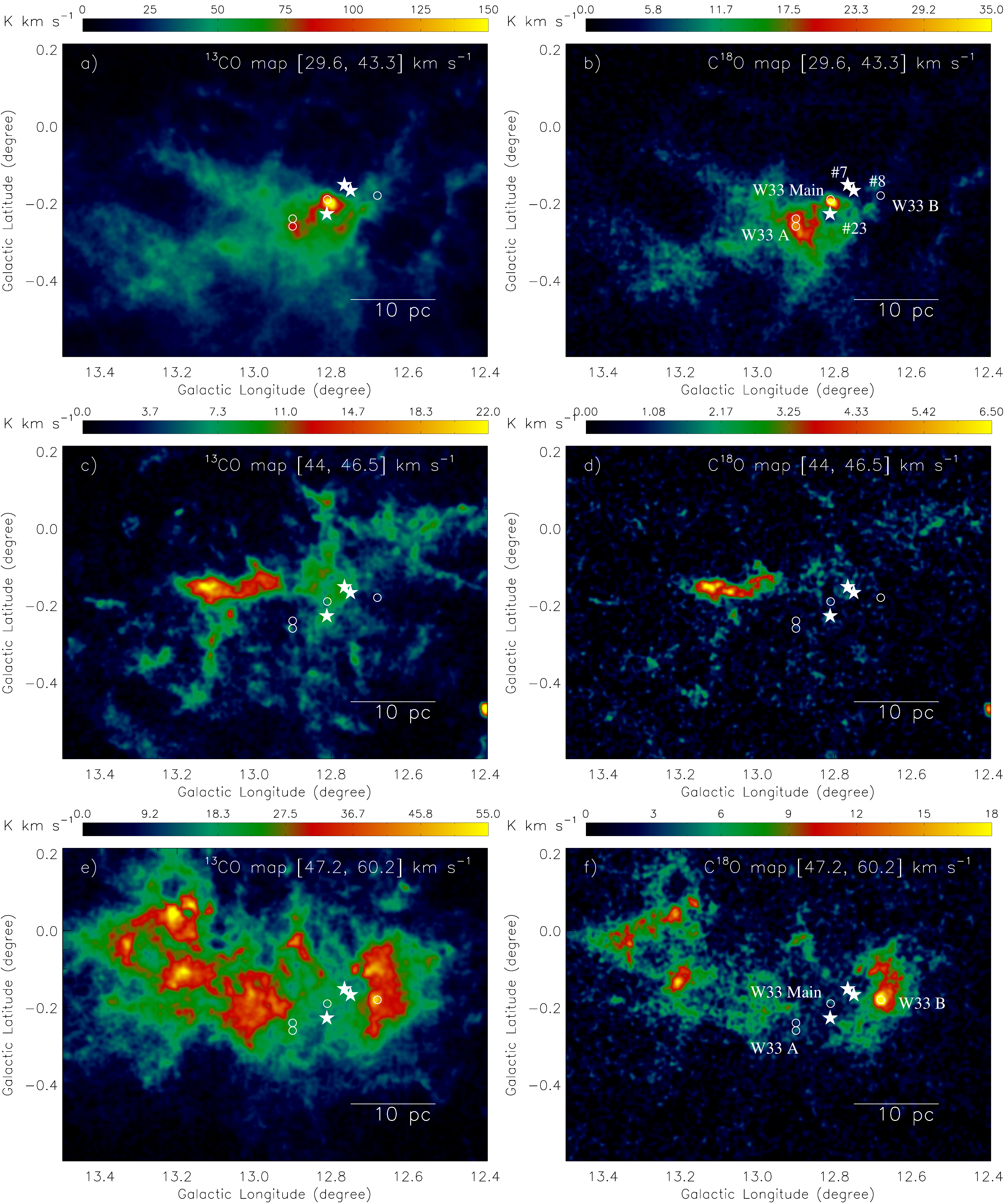}
\caption{a) FUGIN $^{13}$CO(J =1$-$0) map of intensity integrated from 29.6 to 43.3 km s$^{-1}$. 
b) FUGIN C$^{18}$O(J =1$-$0) map at [29.6, 43.3] km s$^{-1}$. 
c) FUGIN $^{13}$CO(J =1$-$0) map of intensity integrated from 44 to 46.5 km s$^{-1}$. 
d) FUGIN C$^{18}$O(J =1$-$0) map at [44, 46.5] km s$^{-1}$. 
e) FUGIN $^{13}$CO(J =1$-$0) map at [47.2, 60.2] km s$^{-1}$. 
f) FUGIN C$^{18}$O(J =1$-$0) map at [47.2, 60.2] km s$^{-1}$. 
In each panel, the other symbols are the same as in Figure~\ref{fig1}b.} 
\label{fig8}
\end{figure*}

Based on the observed different velocity components, Figures~\ref{fig8}a,~\ref{fig8}c, and~\ref{fig8}e show the spatial distribution of $^{13}$CO gas associated with the clouds 
at [29.6, 43.3], [44, 46.5], and [47.2, 60.2] km s$^{-1}$, respectively. 
In Figures~\ref{fig8}b,~\ref{fig8}d, and~\ref{fig8}f, we present the distribution of C$^{18}$O gas associated with the clouds at [29.6, 43.3], [44, 46.5], and [47.2, 60.2] km s$^{-1}$, respectively. In each panel of Figure~\ref{fig8}, the positions of W33 A, W33 B, and W33 Main are also highlighted by open circles, while the positions of three O4-7I stars \citep[see IDs \#7, 8, and 23 in][]{messineo15} are shown by filled stars (in cyan). 
The sites W33 A and  W33 Main are seen with the highest intensities in the $^{13}$CO and C$^{18}$O maps at [29.6, 43.3] km s$^{-1}$, while the corresponding areas are seen with low 
intensities in the $^{13}$CO and C$^{18}$O maps at [47.2, 60.2] km s$^{-1}$. 
One can find a very weak $^{13}$CO and C$^{18}$O emission toward W33 B in the cloud at [29.6, 43.3] km s$^{-1}$, while the strong molecular emission is traced toward W33 B in the cloud 
at [47.2, 60.2] km s$^{-1}$. In the direction of the filamentary structures and the W33 complex, some areas with compact $^{13}$CO and C$^{18}$O emission are also 
seen in map at [44, 46.5] km s$^{-1}$, which is an intermediate velocity range between other two clouds.

\begin{figure*} 
\includegraphics[width=10.2cm]{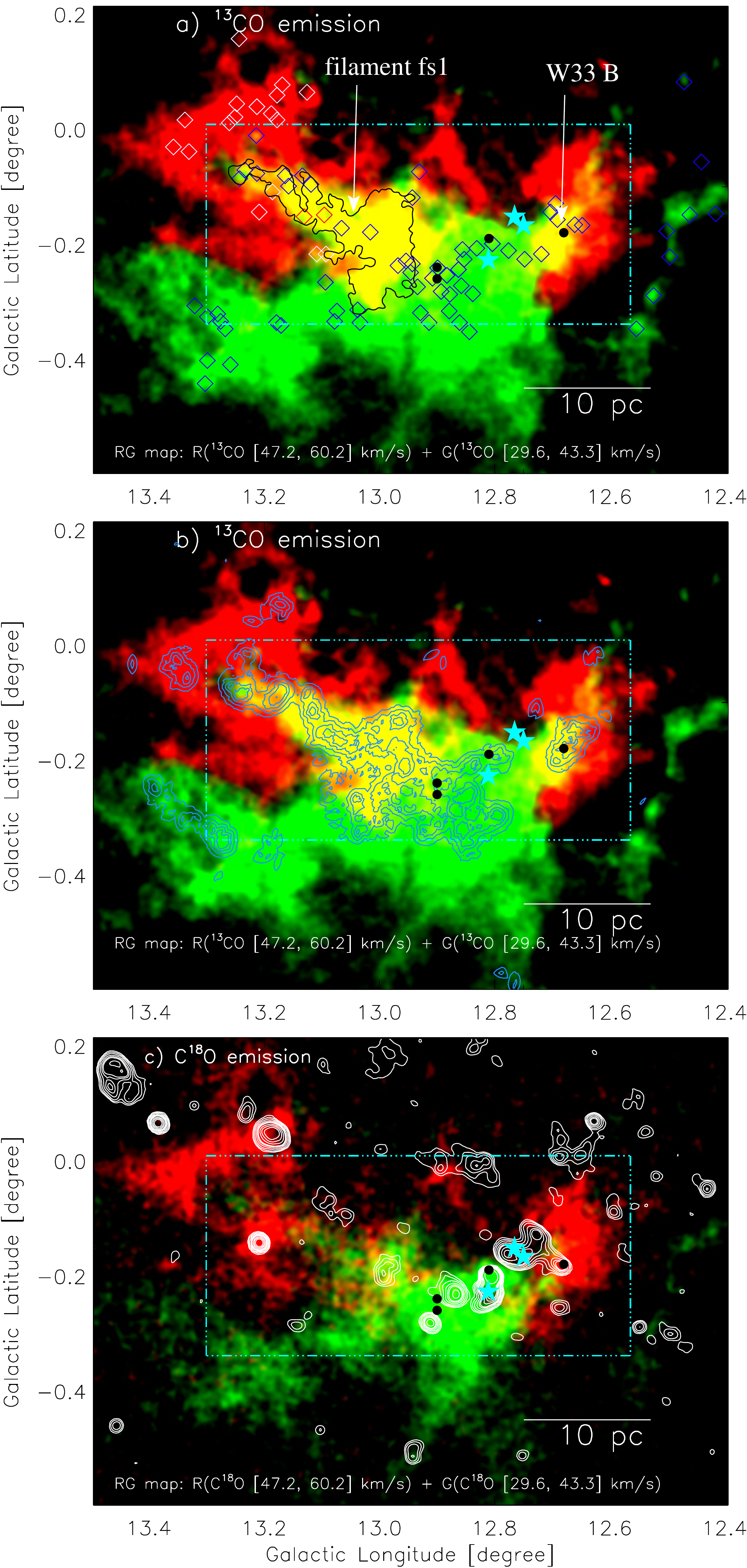}
\caption{a) Two color-composite image produced using the $^{13}$CO maps at [47.2, 60.2] 
and [29.6, 43,3] km s$^{-1}$ in red and green, respectively. The positions of the ATLASGAL 870 $\mu$m dust continuum clumps are marked by diamonds. 
Different colors show the clumps with different velocity ranges (i.e., white diamonds [48.5, 55.4] km s$^{-1}$; red diamonds [44.0, 44.9] km s$^{-1}$; blue 
diamonds [30.5, 40.8] km s$^{-1}$). The filament ``fs1" is indicated by a solid curve (in black), which is traced by the $N(\mathrm H_2)$ contour 
of 3.35 $\times$ 10$^{22}$ cm$^{-2}$ (see Figure~\ref{fig2}b).
b) Overlay of the surface density contours on the color composite map, which is the same as in Figure~\ref{fig9}a.
The surface density contours (in cyan) of Class~I YSOs are shown with the levels of 1.2, 2, 3.5, and 6.5 YSOs/pc$^{2}$. 
c) Overlay of the NVSS radio continuum contours (in white) on the color composite map, which is the same as in Figure~\ref{fig9}a but for C$^{18}$O.
The NVSS radio continuum contours are the same as in Figure~\ref{fig1}b. In each panel, the other symbols are the same as in Figure~\ref{fig1}b.} 
\label{fig9}
\end{figure*}

In Figure~\ref{fig9}a, we show a color-composite image of our selected area with the $^{13}$CO maps at [29.6, 43.3] and [47.2, 60.2] km s$^{-1}$ in red and green, respectively. 
The filamentary feature ``fs1" observed in the {\it Herschel} column density map and the positions of the ATLASGAL clumps are also highlighted in Figure~\ref{fig9}a.   
Figure~\ref{fig9}b displays the overlay of the surface density of Class~I YSOs on the color-composite image produced using the $^{13}$CO maps 
at [29.6, 43.3] and [47.2, 60.2] km s$^{-1}$. A majority of the Class~I YSOs are found toward the common zones of the two cloud components 
(see areas toward the filamentary structures and the W33 complex). 
In Figure~\ref{fig9}c, we present the overlay of the NVSS 1.4 GHz radio continuum contours on the color-composite image produced using the C$^{18}$O maps 
at [29.6, 43.3] and [47.2, 60.2] km s$^{-1}$. Both the color-composite images allow us to infer the overlapping zones of the two clouds at [47.2, 60.2] and [29.6, 43.3] km s$^{-1}$, where the early phase of SF, dust clumps as well as H\,{\sc ii} regions are evident. Furthermore, in the color-composite images, a spatial fit of the intensity-depression region and the intensity-enhancement region is observed toward the W33 complex (see also Figure~\ref{fig10}). 
In the direction of the W33 complex, an intensity-enhancement is found in the molecular intensity maps at [29.6, 43.3] km s$^{-1}$ (see Figures~\ref{fig10}a and~\ref{fig10}b), while an intensity-depression region or cavity-like feature is observed in the molecular intensity maps at [47.2, 60.2] km s$^{-1}$ (see Figures~\ref{fig10}c and~\ref{fig10}d). 
The locations of  W33 A (G012.90$-$0.24, G012.90$-$0.26), W33 Main (G012.81$-$0.19), and three O4-7I stars \citep[see IDs \#7, 8, and 23 in][]{messineo15} are seen toward the areas of the spatial fit of the cavity and the intensity-enhancement region. However, the filament ``fs1" and W33 B are found toward the common zones of the two clouds.

Overall, the study of the molecular line data shows the spatial and velocity connections of two cloud components in the direction of the selected physical system. We have discussed these results in more details in Section~\ref{sec:disc}. 
\section{Discussion}
\label{sec:disc}
\subsection{Physical environment around W33}
As highlighted earlier, previously published results were mainly focused to the 
H\,{\sc ii} regions associated with the W33 complex (area $\sim$15 pc $\times$ 15 pc; see a broken box in Figures~\ref{fig1}a and~\ref{fig1}b). 
The large-scale environment around the W33 complex is not yet studied and explored. 
Such study can enable us to obtain new insights in the target site and to infer different physical mechanism(s) than earlier studies on small-scale environment.
The present paper deals with an extended physical system ($\sim$50 pc $\times$ 37 pc), which is investigated by the distribution of 88 ATLASGAL clumps located at a distance of $\sim$2.6 kpc. 
It is also confirmed by the distribution of molecular gas at [29.6, 60.2] km s$^{-1}$, and at least three velocity components (around 35, 45, and 53 km s$^{-1}$) are 
identified in the direction of our selected physical system (see Section~\ref{sec:coem}). The physical system is found to host the embedded filaments (i.e., fs1 and fs2) and several H\,{\sc ii} regions. 
The filaments are not associated with any radio continuum emission, and exhibit T$_\mathrm{d}$ of $\sim$17--19~K. 
In our selected system, a relatively warm dust emission (T$_\mathrm{d}$ $\sim$ 21--29~K) is also found toward the H\,{\sc ii} regions excited by OB-type stars (see Table~\ref{tab2}). 
These results together show the existence of two distinct environments in the selected physical system, which differ substantially in their dust temperatures (see Section~\ref{sec:hermap}). 
The elongated filament (``fs1"; length $\sim$17 pc; T$_\mathrm{d}$ $\sim$19~K) is characterized as a thermally supercritical filament, and appears to be on the verge of collapse (see Section~\ref{sec:hermap}).
Based on the distribution of Class~I YSOs \citep[mean age $\sim$0.44 Myr;][]{evans09}, the early phase of SF activities is investigated toward the filaments and the H\,{\sc ii} regions (age $\sim$0.35 -- 1.0 Myr) in the system (see Section~\ref{sec:sf}). 
\subsection{Star Formation Scenarios} 
Considering the presence of multiple velocity components and massive O-type stars, one can explore the applicability of the triggered SF scenarios concerning 
the expansion of H\,{\sc ii} regions \citep[e.g.,][]{elmegreen77,bertoldi89,whitworth94,lefloch94,elmegreen98,deharveng05,dale07,bisbas15,walch15,kim18,haid19} 
and the colliding/converging flows from two velocity components \citep{ballesteros99,heitsch08,vazqez07} or the collision of two clouds \citep{elmegreen98}.
Massive OB-stars are the important sources of mechanical and radiative energy, and can impact their surroundings through their 
intense energetic feedback (i.e., stellar wind, ionized emission, and radiation pressure). However, the stellar feedback cannot explain the existence 
of multiple velocity components in the selected physical system \citep[see also][]{kohno18}. 

Concerning the ``cloud cloud collision" (CCC) process, \citet{habe92} studied numerical simulations of a head-on collision of two non-identical clouds. 
The simulations develop gravitationally unstable cores/clumps at the interface of the clouds due to the effect of their compression. 
In the CCC process, young stellar clusters and massive stars can be formed at the intersection of two molecular clouds or the shock-compressed interface layer 
\citep[e.g.,][and references therein]{habe92,anathpindika10,inoue13,takahira14,haworth15a,haworth15b,torii17,bisbas17}. 
In the CCC site, one may observationally find the spatial and velocity connections of two molecular 
clouds \citep[e.g.,][]{torii17,dewangan17x,dewangan17xx,dewangan17b,dewangan18b,dewangan19a}.
In the position-velocity map, one may also trace bridging features between two molecular clouds, 
which enable us to infer the association of the two clouds \citep[e.g.,][]{haworth15a,haworth15b}.
It may also hint the presence of a compressed layer of gas due to two colliding clouds/flows \citep[e.g.,][]{haworth15a,haworth15b,torii17}.
Furthermore, one may also observe a complementary spatial distribution of two clouds in the CCC site 
\citep[e.g.,][]{torii17,fukui18,dewangan18x,dewangan19a}, which is related to the spatial fit of ``Key/intensity-enhancement" 
and ``Cavity/Keyhole/intensity-depression" features.

In the direction of the W33 complex (area $\sim$15 pc $\times$ 15 pc; see a broken box in Figures~\ref{fig1}a and~\ref{fig1}b), \citet{kohno18} proposed the CCC model 
to explain the formation of massive stars. They found complementary distributions of two clouds (at 35 and 58 km s$^{-1}$) around the W33 complex. 
These authors also computed the timescale of the collision between these two clouds (i.e., 0.7--1.0 Myr), which was found to be older than the dynamical timescales of the 
outflows in W33~Main and W33~A \citep[i.e., $\sim$10$^{4}$ yr;][]{kohno18}. 
They concluded that the sources in the W33 complex (i.e., W33~Main, W33~A, W33~B, W33~Main1, W33~A1, and W33~B1) were formed in the dense layer produced via the collision event. 
One can note that previously, no attempt was made to examine the wide-area around the W33 complex, which has been carefully carried out in the present paper. 

In the present work, we have detected three velocity components around 35, 45, and 53 
km s$^{-1}$ toward our selected physical system. \citet{kohno18} reported the cloud at 58 km s$^{-1}$ as the third velocity component using the NANTEN2 $^{12}$CO (J = 1--0) emission. It appears that the cloud at 53 km s$^{-1}$ reported in this paper is a part of the cloud component at 58 km s$^{-1}$ identified by \citet{kohno18} (see Figure~2b in their paper). 
In the direction of the W33 complex and the filamentary 
structures, in the velocity space, the two clouds around 35 and 53 km s$^{-1}$ are linked by lower intensity intermediate velocity emission, which may trace bridge-like features around 45 km s$^{-1}$ (see broken curves around {\it l} = 12$\degr$.8 and {\it l} =13$\degr$.1 in Figure~\ref{fig7}a). 
An analysis of the spatial distribution of molecular gas associated with two clouds around 35 and 53 km s$^{-1}$ 
shows the overlapping areas toward the filament ``fs1" and W33~B (see arrows in Figure~\ref{fig9}a).  
These overlapping areas are ablaze with SF activities including massive stars (see Figures~\ref{fig9}b and~\ref{fig9}c), where we also find materials with high column densities. 

\begin{figure*} 
\includegraphics[width=19.2cm]{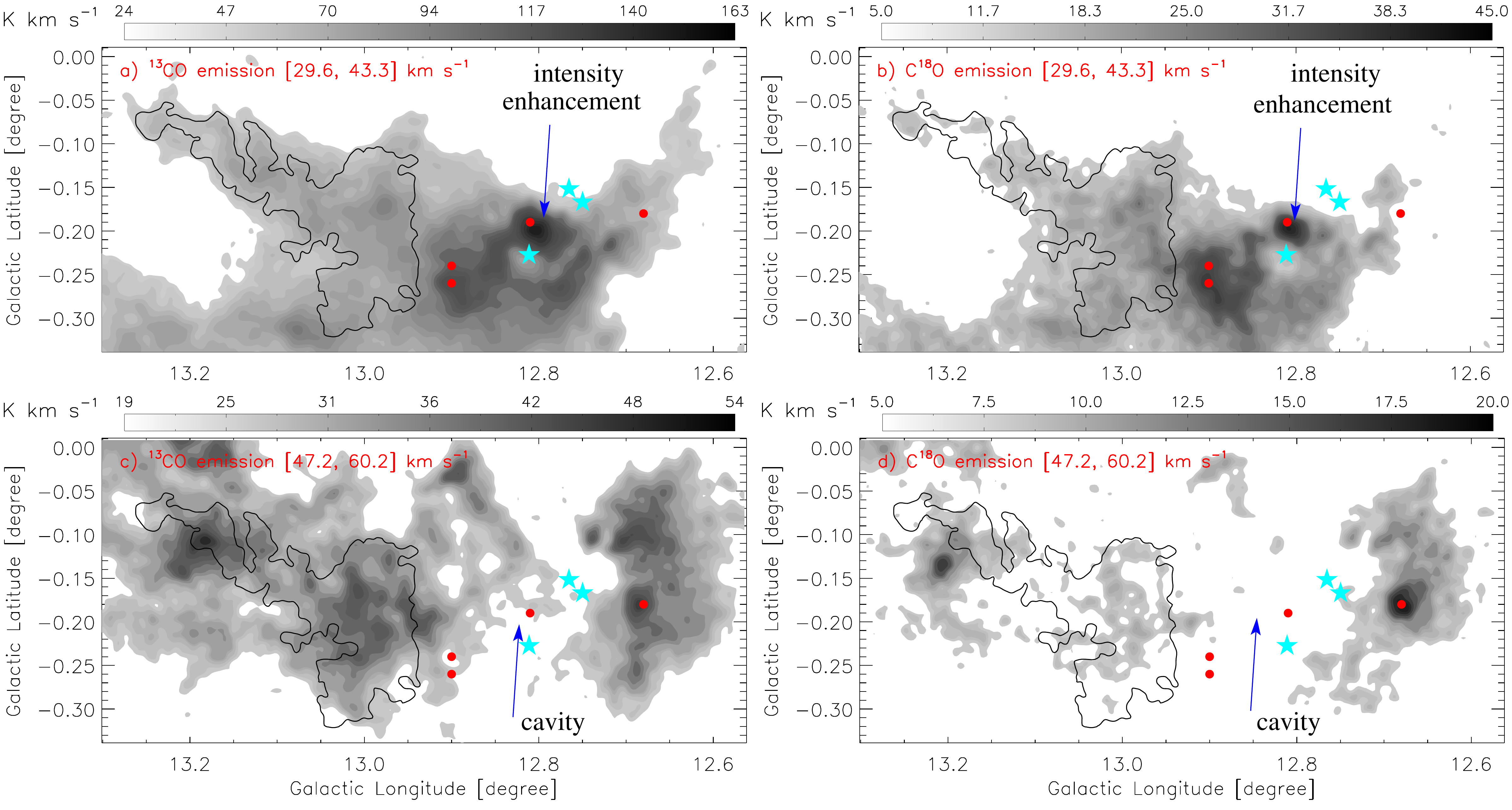}
\includegraphics[width=7.2cm]{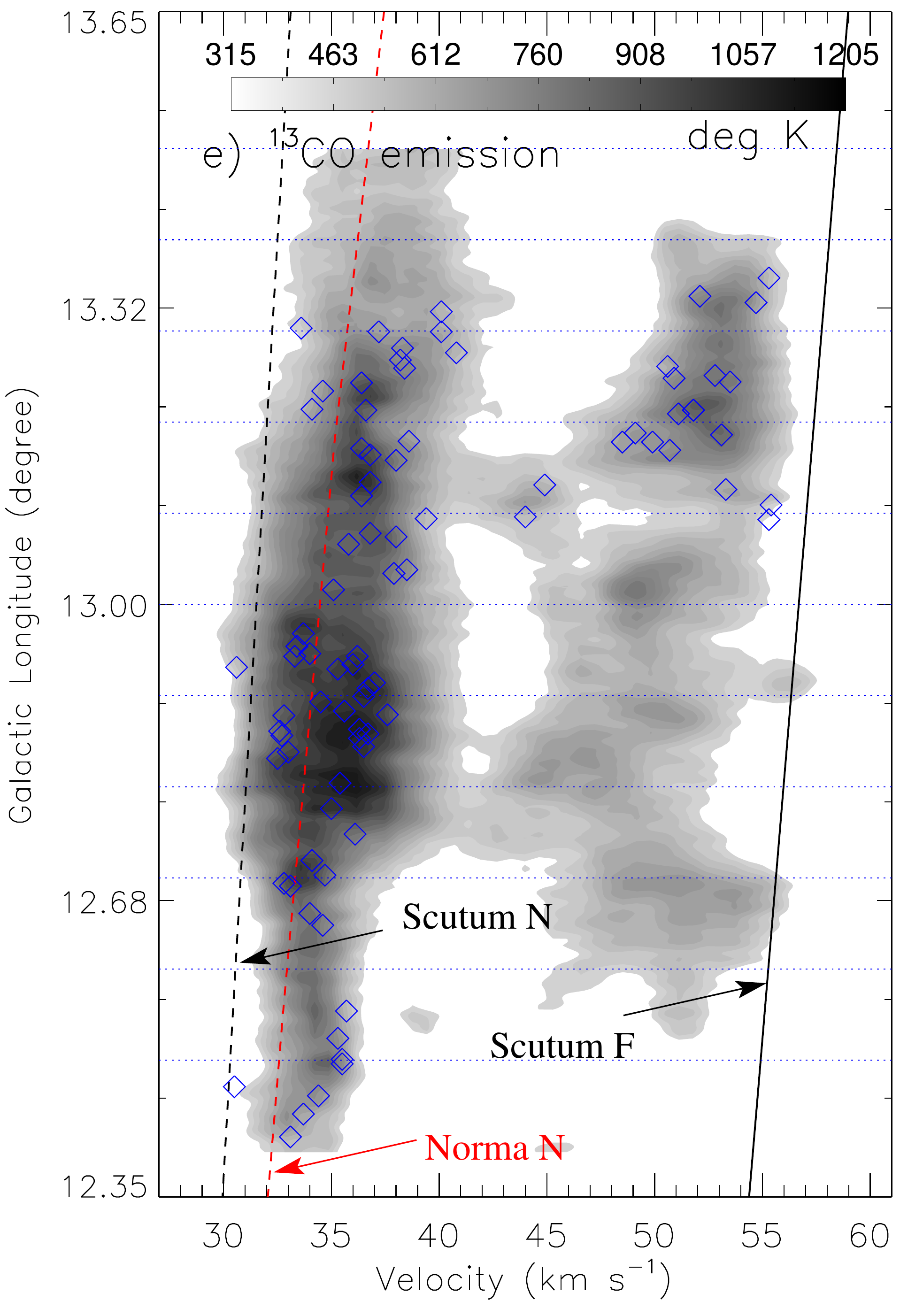}
\caption{Distribution of molecular gas toward an area as highlighted by a broken box in Figure~\ref{fig9}a. 
a) FUGIN $^{13}$CO(J =1$-$0) map of intensity integrated from 29.6 to 43.3 km s$^{-1}$. 
b) FUGIN C$^{18}$O(J =1$-$0) map at [29.6, 43.3] km s$^{-1}$. 
c) FUGIN $^{13}$CO(J =1$-$0) map at [47.2, 60.2] km s$^{-1}$. 
d) FUGIN C$^{18}$O(J =1$-$0) map of intensity integrated from 47.2 to 60.2 km s$^{-1}$. 
f) The panel exhibits the overlay of the Scutum and Norma arms \citep[from][]{reid16} on the longitude-velocity map of $^{13}$CO.
The near and far sides of the arms are marked by broken and solid curves, respectively. 
The radial velocity of each ATLASGAL clump against its longitude is also presented in the plot (see diamonds in Figure~\ref{fig1}a). 
In the panels ``a--d", the other symbols are the same as in Figure~\ref{fig1}b. 
In the panels ``a--d", the filament ``fs1" is indicated by a solid curve (in black), which is traced by the $N(\mathrm H_2)$ contour of 
3.35 $\times$ 10$^{22}$ cm$^{-2}$ (see Figure~\ref{fig2}b).} 
\label{fig10}
\end{figure*}

Additionally, a complementary distribution of two clouds (around 35 and 53 km s$^{-1}$) is evident toward the W33 complex (see Figures~\ref{fig10}a \&~\ref{fig10}c, and~\ref{fig10}b \&~\ref{fig10}d), where the objects W33~Main, W33~A, and three massive O-type stars (i.e., \#7, \#8, and \#23) are located. 
As mentioned earlier, the complementary distribution of clouds was also reported by \citet{kohno18}. 
The key-like or intensity-enhancement feature is traced in the cloud at [29.6, 43.3] km s$^{-1}$ (or around 35 km s$^{-1}$), while the key-hole or cavity-like feature is evident 
in the cloud at [47.2, 60.2] km s$^{-1}$ (or around 53 km s$^{-1}$). However, we do not find any signature of complementary distribution of two clouds in the direction of the filament ``fs1" and W33~B, but these objects are seen toward the common zones of the two clouds as discussed above. Overall, these results suggest the interaction of the two cloud components around 35 and 53 km s$^{-1}$ toward our selected physical system, which is associated with the early phase of SF activities and hosts H\,{\sc ii} regions powered by massive OB stars. Note that our findings can be considered as an extension of the published work of \citet{kohno18}, but for a larger region or area (i.e., $\sim$50 pc $\times$ 37 pc). 
They reported the timescale of the collision to be 0.7--1.0 Myr.

In Figure~\ref{fig10}e, we display the overlay of the Scutum and Norma arms \citep[from][]{reid16} on the longitude-velocity map of $^{13}$CO. 
The radial velocities of 88 ATLASGAL clumps are also shown in Figure~\ref{fig10}e.
Figure~\ref{fig10}e also provides the information of the near side of the Norma and Scutum arms as well as the far side of the Scutum arm. 
Previously, based on the parallax measurements of maser sources, it was reported that the W33 complex is located toward the Scutum arm \citep[see Figure~3 in][]{sato14}. 
Considering the distribution of the molecular gas and the ATLASGAL clumps in the velocity space, one may examine the interaction or collision of the gas clouds associated with the Scutum arm.

Earlier, using high resolution continuum and line data from the Submillimeter Array (SMA) and the Very Large Array (VLA) facilities, 
\citet{galvan10} studied the SF activity toward W33~A, and interprated observed results in W33~A through a triggered SF scenario by filamentary convergent 
gas flows from two different velocity components \citep[see also][]{inoue18}.  
In this scenario, one can expect the origin of molecular clouds via the convergence of streams of neutral gas. 
With time, the merging/converging/collision of the molecular clouds, filaments of molecular gas is expected, which produces the birth of cores and stars \citep{hunter86,ballesteros99,vazqez07,heitsch08,inoue18}.  

Previously, in the case of Orion Nebula Cluster \citep{fukui18} and RCW 38 \citep{fukui16}, it was suggested that low-mass cluster members were formed before the onset of the collision event. However, in our selected target site, Class~I YSOs (i.e., age $\sim$0.44 Myr) and H\,{\sc ii} regions (age $\sim$0.35 -- 0.85 Myr) appear relatively younger than the collision timescale (i.e., 0.7--1.0 Myr) as reported by \citet{kohno18}. Additionally, it was also pointed out that colliding gas flows may also lead to the birth of low-mass stars in regions of compressed gas and dust \citep{hunter86}.

Taking into account the detection of Class~I YSOs (i.e., age $\sim$0.44 Myr) and H\,{\sc ii} regions (age $\sim$0.35 -- 0.85 Myr) toward the common zones of the two cloud components 
around 35 and 53 km s$^{-1}$, the SF history seems to be explained by a triggered SF scenario by converging/colliding flows from two velocity components in our selected system. 
One can also keep in mind that the site G013.210$-$0.144 and the MIR bubbles (N10 and N11) appear away from the overlapping areas of the clouds. Hence, SF activities toward these objects are unlikely induced by the converging flows. 
\section{Summary and Conclusions}
\label{sec:conc}
In this paper, to understand SF mechanisms, we have investigated a large-scale physical system hosting massive star-forming site W33 using a multi-wavelength approach. 
The present work is benefited with the analysis of the existing large-scale survey data at different wavelengths. 
The important results of this work are presented below.\\
$\bullet$ A spatial spread of 88 ATLASGAL 870 $\mu$m dust 
clumps at d $\sim$2.6 kpc and the distribution of molecular gas at [29.6, 60.2] km s$^{-1}$ are utilized to select the extended physical system ($\sim$50 pc $\times$ 37 pc).\\
$\bullet$ The information concerning the location and the radial velocity indicates our selected physical system 
in the direction of the Scutum and Norma arms.\\
$\bullet$ Two distinct environments are investigated in the physical system, which are 
embedded filaments (with T$_{d}$ $\sim$17--19~K) and H\,{\sc ii} regions powered by OB stars (with T$_{d}$ $\sim$ 21--29~K; age $\sim$0.3--1.0 Myr). 
The filaments are not associated with any ionized emission.\\
$\bullet$ An elongated filament (``fs1"; length $\sim$17 pc) is found to be a thermally supercritical filament.\\
$\bullet$ Based on the distribution of Class~I YSOs \citep[mean age $\sim$0.44 Myr;][]{evans09}, the early stage of SF activities is observed toward the filaments and H\,{\sc ii} regions in the cloud.\\
$\bullet$ The FUGIN $^{13}$CO and C$^{18}$O line data trace three velocity components (around 35, 45, and 53 km s$^{-1}$) in the direction of the system.\\
$\bullet$ The analysis of the molecular line data confirms the spatial and velocity connections of cloud components around 35 and 53 km s$^{-1}$. \\
$\bullet$ The positions of W33~Main, W33~A and O4-7I stars are found toward a complementary distribution 
of cloud components around 35 and 53 km s$^{-1}$.\\
$\bullet$ The spatial distribution of the clouds around 35 and 53 km s$^{-1}$ reveals overlapping zones toward the filament ``fs1" and an object W33~B, where a majority of the ATLASGAL clumps and noticeable Class~I objects are depicted. \\\\

We conclude that the observed SF signposts seem to be explained well by a triggered SF scenario by converging/colliding flows from two velocity components around 35 and 53 km s$^{-1}$ in the physical system hosting two distinct environments (i.e., H\,{\sc ii} regions and filamentary structures).
\section*{Acknowledgments}
We thank the anonymous reviewer for several useful comments and suggestions. 
The research work at Physical Research Laboratory is funded by the Department of Space, Government of India. 
This work is based [in part] on observations made with the {\it Spitzer} Space Telescope, which is operated by the Jet Propulsion Laboratory, California Institute of Technology under a contract with NASA. 
This publication makes use of data from FUGIN, FOREST Unbiased Galactic plane Imaging survey with the Nobeyama 45-m telescope, a legacy project in the Nobeyama 45-m radio telescope. 
TB is supported by the National Key Research and Development Program of China through grant 2017YFA0402702. 
TB also acknowledges support from the China Postdoctoral Science Foundation through grant 2018M631241. 
DKO acknowledges the support of the Department of Atomic Energy, Government of India, under project No. 12-R\&D-TFR-5.02-0200.
\begin{table*}
\setlength{\tabcolsep}{0.1in}
\centering
\caption{Physical parameters of radio clumps traced in the NVSS 1.4 GHz continuum map (see Figure~\ref{fig1}c). 
Table tabulates ID, Galactic coordinates ({\it l}, {\it b}), deconvolved effective radius of the H\,{\sc ii} region ($R_\mathrm{HII}$), total flux ($S{_\nu}$), 
Lyman continuum photons ($\log{N_\mathrm{uv}}$), dynamical age ($t_\mathrm{dyn}$), and radio spectral type.} 
\label{tab2}
\begin{tabular}{lcccccccccccccr}
\hline 
  ID  &  {\it l}     &  {\it b}    &  $R_\mathrm{HII}$   & $S{_\nu}$  & $\log{N_\mathrm{uv}}$  & $t_\mathrm{dyn}$ (Myr) &              Spectral Type   \\  
      & (degree) & (degree) &  (pc)               &  (Jy)      &  (s$^{-1}$)           &  for $n_{0}$ = 10$^{4}$ cm$^{-3}$                    &     (V)  \\  
\hline
\hline 
  c1  &   12.731  &   $-$0.130 &	  1.31 &   688.81  &  47.56  &   0.79  & B0.5-B0 \\
  c2  &   12.686  &   $-$0.188 &	  1.08 &   361.55  &  47.28  &   0.66  & B0.5-B0 \\ 
  c3  &   12.781  &   $-$0.167 &	  1.39 &  1477.56  &  47.89  &   0.72  & B0-O9.5 \\ 
  c4  &   12.811  &   $-$0.205 &	  1.53 & 11370.76  &  48.78  &   0.51  & O7.5-O7 \\
  c5  &   12.869  &   $-$0.234 &	  1.34 &   653.54  &  47.54  &   0.83  & B0.5-B0 \\ 
  c6  &   12.915  &   $-$0.284 &	  0.81 &   361.55  &  47.28  &   0.39  & B0.5-B0 \\ 
  c7  &   12.998  &   $-$0.201 &	  1.12 &   163.63  &  46.94  &   0.86  & B0.5-B0 \\
  c8  &   13.215  &   $-$0.147 &	  0.87 &   888.69  &  47.67  &   0.36  & B0-O9.5 \\ 
  c9  &   13.090  &   $-$0.072 &	  1.16 &    92.10  &  46.69  &   1.05  & B0.5-B0 \\
  c10 &   13.190  &      0.045 &	  1.43 &  2698.41  &  48.16  &   0.65  & O9.5-O9 \\ 
  c11 &   13.236  &      0.074 &	  0.80 &    76.43  &  46.61  &   0.57  & B0.5-B0 \\
\hline          
\end{tabular}
\end{table*}

\end{document}